\documentclass{Rinton-P9x6}

\def\npb#1#2#3{    {\it Nucl. Phys. }{\bf B #1} (#2) #3}

\def\plb#1#2#3{    {\it Phys. Lett. }{\bf B #1} (#2) #3}
\def\prd#1#2#3{    {\it Phys. Rev. }{\bf D #1} (#2) #3}

\def\prl#1#2#3{    {\it Phys. Rev. Lett. }{\bf #1} (#2) #3}

\def\lsim{\raise0.3ex\hbox{$\;<$\kern-0.75em\raise-1.1ex\hbox{$\sim\;$}}}
\def\gsim{\raise0.3ex\hbox{$\;>$\kern-0.75em\raise-1.1ex\hbox{$\sim\;$}}}

\def\bmat{\left(\begin{array}}
\def\emat{\end{array}\right)}
\def\Ibanez{Iba\~{n}ez~}   
\def\Munoz{Mu\~{n}oz~}
\begin{document}

\title{Lepton flavor violation in supersymmetric models with non--degenerate
$A$--terms}

\author{D.F. Carvalho and M.E. G\'omez}

\address{ CFIF, Departamento de F\'{\i}sica,
Instituto Superior T\'ecnico, Av. Rovisco
Pais,\\[-0.3em] 1049-001 Lisboa, Portugal}

\author{S. Khalil}

\address{Centre for Theoretical Physics, University of Sussex,
Brighton BN1 9QJ, U.K.\\
Ain Shams University, Faculty of Science, Cairo, 11566, Egypt.}


\maketitle
\maketitle

\abstracts{We analyze the lepton--flavor violation in supersymmetric 
models with non--universal soft breaking terms derived from strings.
We show that the non--universality of the scalar masses
enhances the branching ratios more significantly than the
non--universality of the $A$--terms.}

\section{Introduction}
Non--universal $A$--terms provide an interesting framework to
enhance the supersymmetric contributions to CP violation
effects~\cite{khalil1,khalil2,vives,gordy2}.
It was shown that the flavor structure of the $A$--terms is
crucial for enhancing the SUSY contributions to CP violation
effects, and for generating the experimentally observed $\varepsilon$
and $\varepsilon'/\varepsilon$. These models also
predict a CP asymmetry in $B \to X_s \gamma$ decay much
larger than the one predicted by SM and the one obtained for
a wide region of the parameter space in
minimal supergravity scenarios~\cite{bailin}.
It has also been noted
that the analysis of $b \to s \gamma$ does not severely constrain
the models under consideration~\cite{emidio}. However,
the non--universality of the
$A$--terms in this class of models is always associated with
non--universal  scalar masses. In this case a simple
non--diagonal Yukawa texture predicting lepton masses will induce
two sources of
LFV: one due to the flavor structure of the $A_l$--terms which
prohibits the simultaneous diagonalization of the lepton Yukawa matrices
$Y_l$ and the trilinear couplings $(Y_l^A)_{ij}\equiv (A_l)_{ij}
(Y_l)_{ij}$; the other source is due to the non degeneracy of
the scalar masses of the sleptons. Therefore, in the basis where $m_l$ is
diagonal the slepton mass matrix acquires non--diagonal
contributions. We find that in general the second source dominates
over the first in the case of the LFV predictions.

\section{String inspired models with non-degenerate  $A$--terms}   
In this work we consider the class of string inspired model which
has been recently studied in
Refs.~\cite{khalil1,khalil2,vives}. In
this class of models, the trilinear $A$--terms of the soft SUSY
breaking are non--universal. It was shown that this  
non--universality among the $A$--terms plays an important role on
CP violating processes. In particular, it has been shown that
non-degenerate $A$-parameters can generate the experimentally
observed CP violation $\varepsilon$ and $\varepsilon'/\varepsilon$
even with a vanishing $\delta_{\mathrm{CKM}}$.

Here we consider two models for non-degenerate $A$--terms. The
first model (model A) is based on weakly coupled heterotic strings,
where the dilaton and the moduli fields contribute to SUSY
breaking~\cite{ibanez1}. The second model (model B) is based on type
I string theory where the gauge group $SU(3) \times U(1)_Y$ is
originated from the $9$ brane and the gauge group $SU(2)$ is
originated from one of the $5$ branes~\cite{ibanez2}.

\subsection{Model A}
We start with the weakly coupled string-inspired supergravity
theory.
In this class of models, it is assumed that the superpotential of
the dilaton ($S$) and moduli ($T$) fields is
generated by some non-perturbative mechanism and
the $F$-terms of $S$ and $T$ contribute to the SUSY breaking.
Then one can parametrize the $F$-terms as~\cite{ibanez1}  
\be
F^S = \sqrt{3} m_{3/2} (S+S^*) \sin\theta,\hspace{0.75cm} F^T 
=m_{3/2} (T+T^*) \cos\theta .
\ee
Here $m_{3/2}$ is the gravitino mass, $n_i$ is the modular weight
and $\tan \theta$ corresponds to the ratio between the $F$-terms of $S$
and $T$.
In this framework, the soft scalar masses $m_i$ and the gaugino masses
$M_a$ are given by~\cite{ibanez1}
\begin{eqnarray}
m^2_i &=& m^2_{3/2}(1 + n_i \cos^2\theta), \label{scalar}\\ M_a
&=& \sqrt{3} m_{3/2} \sin\theta .\label{gaugino}
\end{eqnarray}
The  $A^{u,d}$-terms are  written as
\begin{eqnarray}
(A^{u,d})_{ij} &=& - \sqrt{3} m_{3/2} \sin\theta- m_{3/2}
\cos\theta (3 + n_i + n_j + n_{H_{u,d}}), \label{trilinear}
\end{eqnarray}
where $n_{i,j,k}$ are the modular weights of the fields
that are coupled by this $A$--term.
If we assign $n_i=-1$ for the third family and $n_i=-2$
for the first and second
families (we also assume that $n_{H_1}=-1$ and $n_{H_2}=-2$) we find the following
texture for the $A$-parameter matrix at the string scale
\begin{equation}
A^{u,d} = \left (
\begin{array}{ccc}
x_{u,d} & x_{u,d} & y_{u,d}\\
x_{u,d} & x_{u,d} & y_{u,d} \\
y_{u,d} & y_{u,d} & z_{u,d}
\end{array}   
\right),
\label{AtermA}
\end{equation}
where
\begin{eqnarray}
x_u&=& m_{3/2}(-\sqrt{3} \sin\theta + 3  \cos\theta),\\ 
x_d&=&y_u= m_{3/2}(-\sqrt{3} \sin\theta + 2  \cos\theta),\\
y_d&=&z_u= m_{3/2}(-\sqrt{3} \sin\theta + \cos\theta),\\
z_d&=&-\sqrt{3}m_{3/2}\sin\theta.
\end{eqnarray}

The non--universality of this model is
parameterized by the angle $\theta$ and the value $\theta =\pi/2$
corresponds to the universal limit for the soft terms. In order to
avoid negative mass squared in the scalar masses we restrict
ourselves to the case with $\cos^2 \theta < 1/2$. Such
restriction on $\theta$ makes the non--universality in the whole
soft SUSY breaking terms very limited. However, as shown in
\cite{khalil1,khalil2}, this small range of variation for the
non--universality is enough to generate sizeable SUSY CP violations
in K system.

\subsection{Model B}
This model is based on type I string theory and like model A, it is a good candidate for
generating sizeable SUSY CP violations. In type I string theory, non--universality
in the scalar masses, $A$--terms and gaugino masses
can be naturally obtained~\cite{ibanez2}. Type I models contain   
either 9 branes and three types of $5_i (i=1,2,3)$ branes or $7_i$
branes and 3 branes. From the phenomenological point of view there
is no difference between these two scenarios. Here we consider the
same model used in Ref.~\cite{vives}, where the gauge group
$SU(3)_C \times U(1)_Y$ is associated with 9 brane while $SU(2)_L$
is associated with $5_1$ brane.

If SUSY breaking is analysed, as in model A,
in terms of the vevs of the dilaton and moduli fields \cite{ibanez2}
\be
F^S = \sqrt{3} m_{3/2} (S+S^*) \sin\theta,\hspace{0.75cm} F^{T_i}
=m_{3/2} (T_i+T_i^*) \Theta_i \cos\theta~,
\ee
where the angle $\theta$ and the parameter $\Theta_i$ with
$\sum_i \left|\Theta_i\right|^2=1$, just parametrize the direction of the
goldstino in the $S$ and $T_i$ fields space .
Within this framework, the gaugino masses are~\cite{ibanez2}
\bea
M_1 &=& M_3 = \sqrt{3} m_{3/2} \sin\theta ,\\ M_2 &=& \sqrt{3}
m_{3/2} \Theta_1 \cos \theta .\label{m2}
\label{gauginoB}
\eea
In this case the quark doublets and the Higgs fields are assigned to
the open string which spans between the $5_1$ and $9$ branes.
While the quark singlets correspond to the open string which starts
and ends on the $9$ brane, such open string includes three sectors
which correspond to the three complex compact dimensions. If we
assign the quark singlets to different sectors we obtain
non--universal $A$--terms. It turns out that in this model the
trilinear couplings $A^u$ and $A^d$ are given
by~\cite{ibanez2,vives}
\begin{equation}
A^u=A^d = \left (
\begin{array}{ccc}
x & y & z\\
x & y & z \\
x & y & z
\end{array}
\right),
\label{AtermB1}
\end{equation}
where
\begin{eqnarray}
x &=& - \sqrt{3} m_{3/2}\left(\sin\theta + (\Theta_1 - \Theta_3) \cos\theta
\right),\\
y &=& - \sqrt{3} m_{3/2}\left(\sin\theta + (\Theta_1 - \Theta_2) \cos\theta
\right),\\
z &=& - \sqrt{3} m_{3/2} \sin\theta.
\label{AtermB2}
\end{eqnarray}
The soft scalar masses for quark-doublets and Higgs fields
$(m^2_L)$, and the quark-singlets $(m^2_{R_i})$ are given by
\bea
m^2_L &=&  m_{3/2}^2 \left( 1- \frac{3}{2} (1-\Theta_1^2) \cos^2
\theta\right) ,\\ m^2_{R_i} &=&  m_{3/2}^2 \left( 1- 3 \Theta_i^2 \cos^2
\theta\right),  
\label{scalarB}
\eea
where $i$ refers to the three families.
For $\Theta_{i} = 1/\sqrt{3}$ the $A$--terms and the
scalar masses are universal while the gaugino masses could be
non--universal. The universal gaugino masses are obtained at
$\theta=\pi/6$.

In models with non-degenerate $A$--terms we have to fix the Yukawa
matrices to completely specify the model.
Here we assume that the Yukawa texture has the following form  
\begin{itemize}
\item Texture I,
\be
Y_{l}=y^{\tau }\left( \begin{array}{ccc}
0 & 5.07\times 10^{-3} & 0\\
5.07\times 10^{-3} & 8.37\times 10^{-2} & 0.4\\
0 & 0.4 & 1
\end{array}\right)
\ee
\item Texture II, 
\be
Y_{l}=y^{\tau }\left( \begin{array}{ccc}
3.3\times 10^{-4} & 1.64\times 10^{-5} & 0\\
1.64\times 10^{-5} & 8.55\times 10^{-2} & 0.4\\
0 & 0.4 & 1
\end{array}\right).
\ee
\end{itemize}   
Both of them lead to the correct prediction for the experimental values of
the lepton masses.

\section{Lepton flavor violation vs. non--universality}

We are now ready to investigate the lepton flavor violation processes 
$\tau \to \mu~ \gamma$ and $\mu \to e~ \gamma$~\cite{daniel}.
Despite lepton flavor is preserved in the standard model (SM), SUSY
models can predict branching ratios for these processes compatible with
present experimental bounds as it has been discussed by J. Casas in this 
conference. The details of the calculations for these ratios in the
context of the models described in section 2 are in Ref.\cite{lfv}.     
As emphasized in section 2, the splitting of the
soft masses in model A increases from $\sin\theta=1$ (which corresponds to 
the universal case) to the limiting case for  $\sin \theta=1/\sqrt{2}$
(below which some  square masses to become negative). Therefore we
consider as representative for this model to present  the
variation of the  branching
ratios with  $\sin\theta$ for fixed values of  $m_{3/2}$ as
shown in Fig. 1. 
\begin{figure}[ht]
\hspace{-0.2cm}
\begin{minipage}[b]{9in}
\epsfig{figure=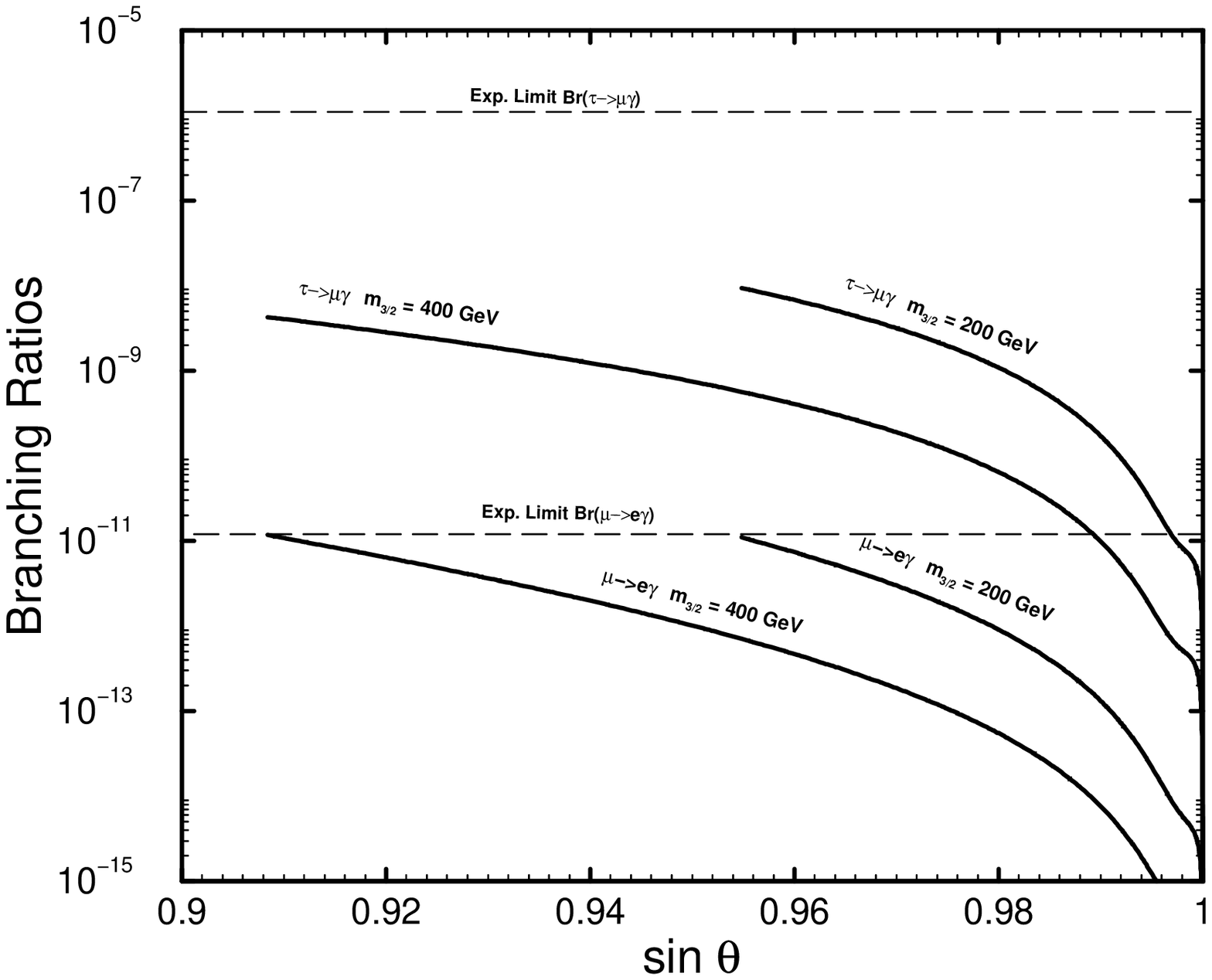,height=2.3in,width=2.2in,angle=0}
\hspace{0.5cm}\epsfig{figure=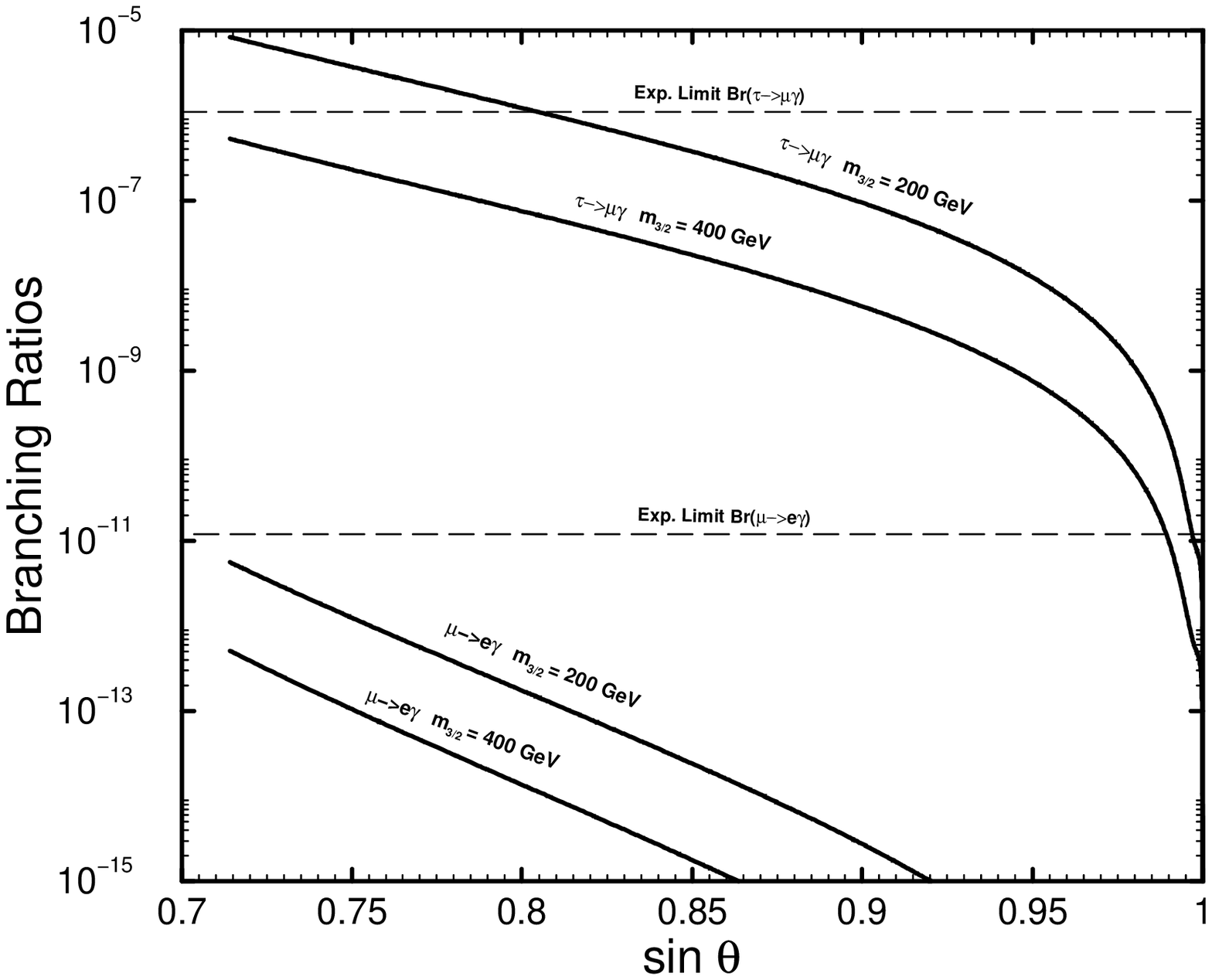,height=2.3in,width=2.2in,angle=0}
\end{minipage}
\medskip
\caption{Branching ratios vs. $\sin\theta$ for model A with
texture I for $Y_l$ (left) and texture II (right) and $\tan\beta=10$.
The values for $m_{3/2}$ are kept constant as shown on the curves.
\label{fig.2}}
\end{figure}

For the value $m_{3/2}=200\ \rm{GeV}$, the mass of lightest neutralino varies
from 100 to 147 GeV, that of
the chargino from 190 to 277 GeV, while the lightest of the staus has masses
of  107 to 233 GeV as $\sin \theta$ ranges from $1/\sqrt{2}$ to 1. Similarly
for $m_{3/2}=400\ \rm{GeV}$ we found $m_{\tilde{\chi}^0}=210-295 \ \rm{GeV}$,
$m_{\tilde{\chi}^+}=400-570 \ \rm{GeV}$,
$m_{\tilde{\tau}_2}=212-470 \ \rm{GeV}$ for the same range of $\sin \theta$.

From this figure we can see how texture I (graphic on the left) allows
small deviations from universality of the soft terms. The experimental 
bound on $BR(\mu\to e \gamma)$ is satisfied only for $\sin \theta> .96$ 
($m_{3/2}=200\ \rm{GeV}$) and for $\sin \theta> .91$
($m_{3/2}=400\ \rm{GeV}$) while for the same range on  $\sin \theta$
the corresponding prediction for $BR(\tau\to \mu \gamma)$ is well below  
the experimental bound. The values of the branching ratios decrease as
we increase $m_{3/2}$ and $\sin\theta$ since this translates into  an
increase of the masses of the supersymmetric particles. In order
to simplify  the presentation of
our results we fix $\tan\beta=10$. However, enlarging the
value of $\tan\beta$ increases the prediction for the branching ratios.

The results obtained using texture II (Fig.~2, graphic on the right)
allow us to start the graph at the lowest value of $\sin
\theta=1/\sqrt{2}$. As it can  be seen, the experimental bounds are
more restrictive for the $\tau \to \mu \gamma$ than for $\mu \to e 
\gamma$ process. 

We turn now to study the lepton flavor violation in model B. 
In this model, the structure of the soft-terms is more
complicated than in the previous model. They depend, in addition
to  $m_{3/2}$ and $\theta$, on the values of the parameters
$\Theta_1$, $\Theta_2$ and $\Theta_3$. However, the flavor structure of the
slepton matrices is simpler, since the soft masses for the left--handed
sleptons  are universal at the GUT scale and the sneutrino mass
matrix remains diagonal under a rotation that diagonalizes
$Y_l$.  Therefore diagram with exchanging sneutrino will not contribute to the
LFV processes. Also we found that a bound of 
$m_{\tilde {\chi}^+}=95 \ \rm{GeV}$ is found to be the most  
severe on the initial conditions. The predictions we obtain with this
model for  $BR(l_j\to l_i \gamma)$ allow us to simplify our
presentation by setting   
$\Theta\equiv\Theta_1=\Theta_2$. In the case of texture I, this  
is justified by the fact that the experimental
bound on $BR(\mu\to e \gamma)$ tolerates a small deviation
of the $\Theta_i$~'s from the common value of $1/\sqrt{3}$,  for   
which the soft masses become universal as shown in 
Fig. 2. 
\begin{figure}[ht]
\hspace*{-0.2cm}  
\epsfig{figure=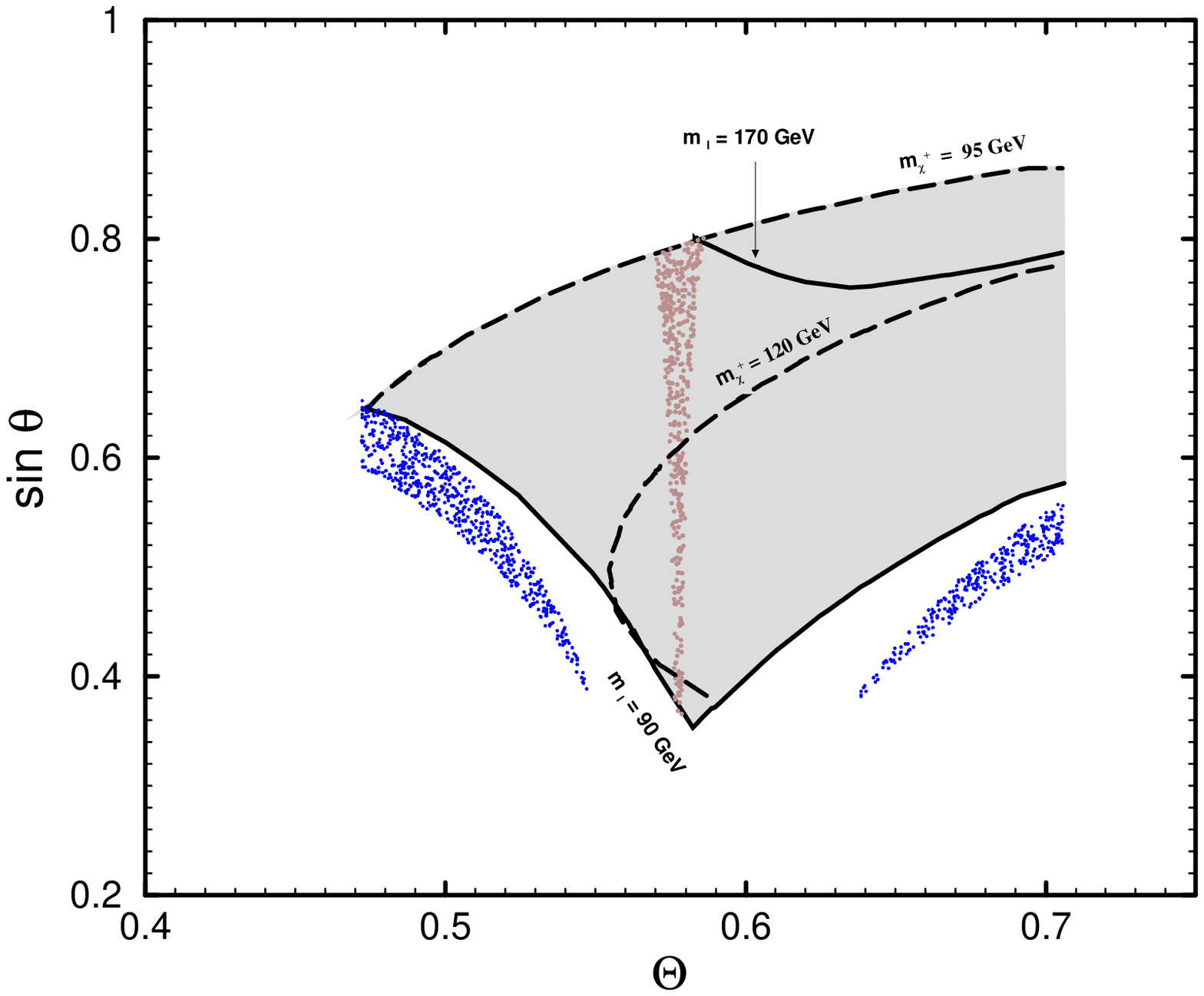,height=2.3in,width=2.2in,angle=0}
\hspace{0.5cm}\epsfig{figure=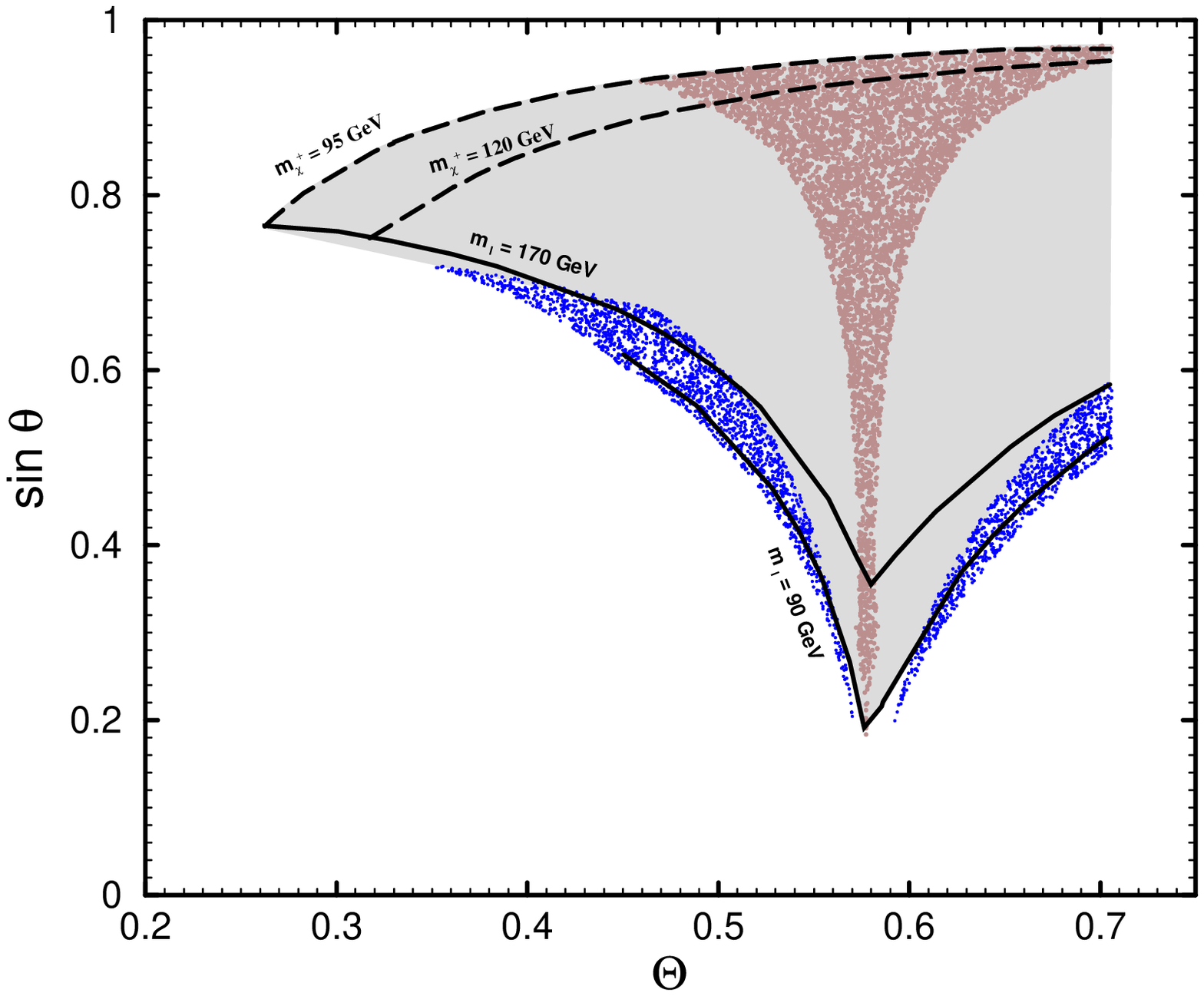,height=2.3in,width=2.2in,angle=0}
\caption{Areas with $BR(\mu\to e \gamma)< 1.2 \times 10^{-11}$
(dotted areas inside the gray contour) in the plane
$\sin\theta-\Theta$ for constant values of
$m_{3/2}=200\ \rm{GeV}$ (left) and
$m_{3/2}=400\ \rm{GeV}$ (right) and $\tan\beta=10$.
The model used corresponds to type I string, with texture I for $Y_l$.
Values of the masses of SUSY particles which bound the parameter
space of the model are as shown in the graphs. The dark dotted areas
correspond to space of parameters such that the LSP is a slepton.
\label{fig.3}}
\end{figure}

For the case of texture II, these predictions are
more tolerant
to a variation of the $\Theta_i$~'s. However, when this texture is
considered,  the
experimental
limit on $BR(\tau\to \mu \gamma)$ is more restrictive
since this bound is particularly
sensitive to the value $\Theta_3$. Therefore, we find that setting also
 $\Theta_1=\Theta_2$ in the analysis of our results with texture II,
we can achieve a clearer presentation without any loss of generality.

Fig.~2 shows the constraint imposed by the current bound on the
$BR(\mu\to e \gamma)$ on the plane ($\sin\theta-\Theta$) for
constant values of $m_{3/2}=200 \ \rm{GeV}$ (left) and
$m_{3/2}=400 \ \rm{GeV}$ (right) when texture I is assumed.
Fig.3 displays  the equivalent for texture II. The light shaded areas
correspond to the space of parameters allowed by the bounds on
the masses of the SUSY particles. The region below the upper
dashed line corresponds to values of
 $m_{\tilde {\chi}^+}> 95 \ \rm{GeV}$,  while the sector above the lower
solid line corresponds to values of the lightest charged scalar
$m_{\tilde {l}}> 90 \ \rm{GeV}$.

\begin{figure}[ht]
\hspace*{-0.2cm}
\epsfig{figure=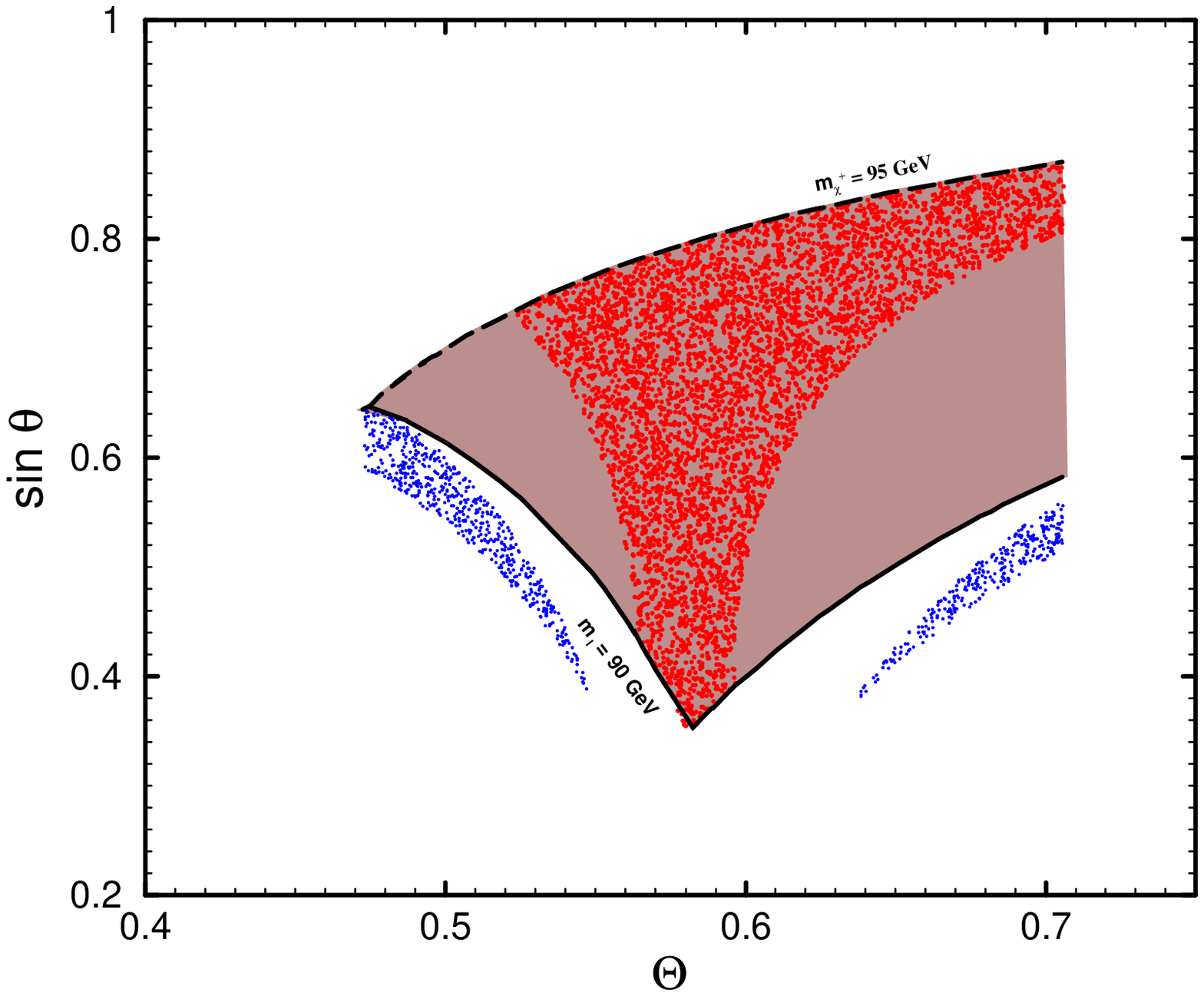,height=2.3in,width=2.2in,angle=0}
\hspace{0.5cm}\epsfig{figure=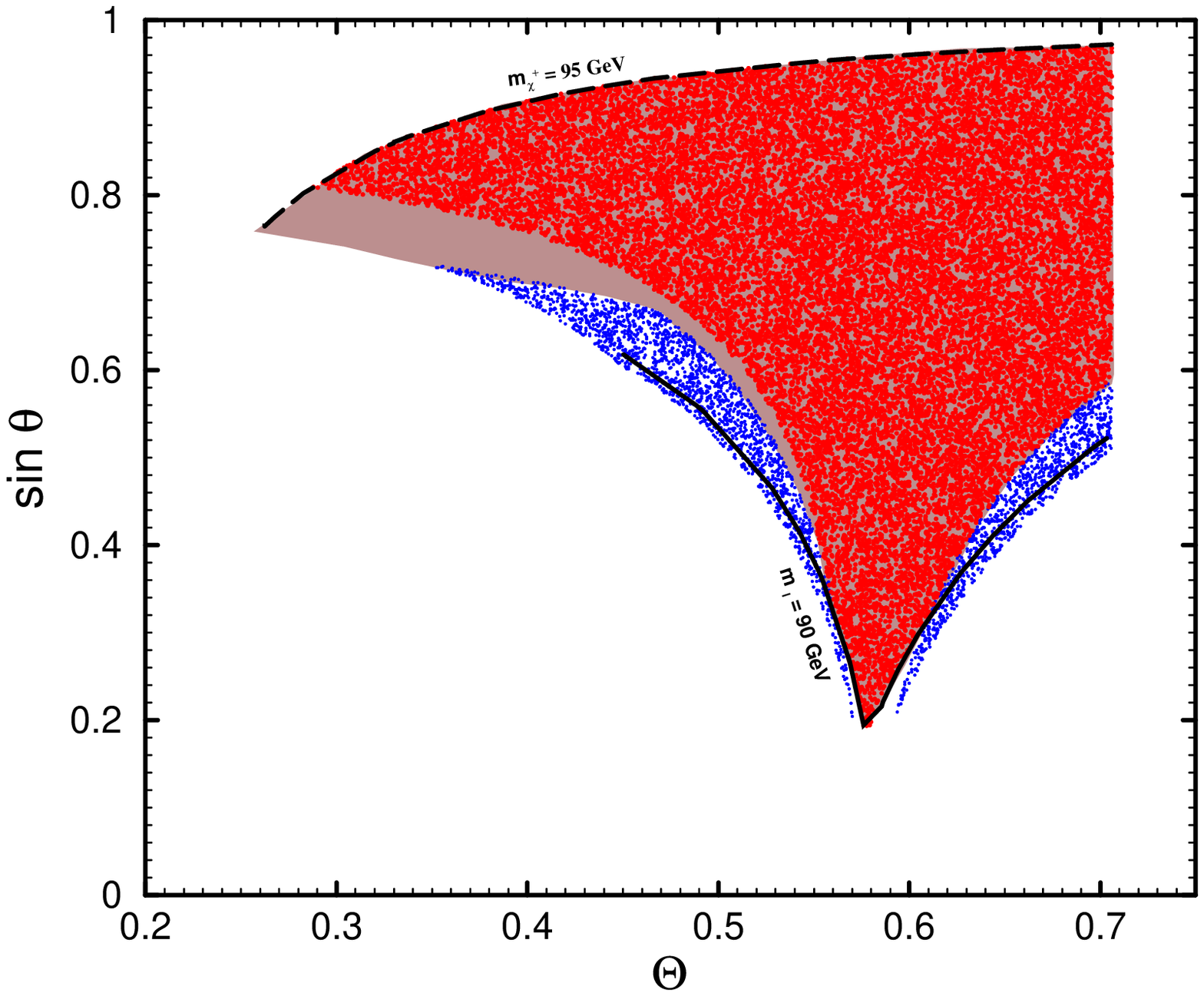,height=2.3in,width=2.2in,angle=0}
\caption{Areas with $BR(\tau\to \mu \gamma)< 1.1 \times 10^{-6}$
(dotted areas inside the gray contour) in the plane
$\sin\theta-\Theta$ for constant values of
$m_{3/2}=200\ \rm{GeV}$ (left) and
$m_{3/2}=400\ \rm{GeV}$ (right) and  $\tan\beta=10$.
The model used corresponds to type I string, with texture II for $Y_l$.
Values of the masses of SUSY particles which bound the parameter
space of the model are as shown in the graphs. The dark dotted areas
correspond to space of parameters such that the LSP is a slepton.}
\end{figure}

The shape of the curve  $m_{\tilde {l}}= 90 \ \rm{GeV}$ in Figs.~2 and 3 is
determined by the initial conditions given above. The lowest
values for these masses corresponds
to  $m_{R_1}=m_{R_2}$ when $1/\sqrt{3}<\Theta<1/\sqrt{2}$, while
for  $\Theta<1/\sqrt{3}$, $m_{R_3}$ is
the lowest value. Therefore the largest component of the lowest
eigenvalue of the charged slepton mass is the $\tilde{e}_R$ or
the $\tilde{\tau}_R$ depending on the ranges of $\Theta$ above. Similar
considerations explain the different shape of
the curves for $m_{\tilde {l}}= 170 \ \rm{GeV}$ (with $m_{3/2}=200 \ \rm{GeV}$
and $m_{3/2}=400 \ \rm{GeV}$).

The darkest dotted areas in  Figs.~2 and 3 represent the sector
of parameters for which the lightest supersymmetric particle (LSP)
is a charged slepton. For $m_{3/2}=200 \ \rm{GeV}$ these areas are
below the bound of
$m_{\tilde {l}}= 90 \ \rm{GeV}$. However for  $m_{3/2}=400 \ \rm{GeV}$, the
cosmological requirement on the LSP to be a neutral particle (lightest
neutralino in our case) imposes a further restriction on the space of
parameters of the model.

Similarly to the results found for the model A,
the assumption of  texture I for $Y_l$ allows a small deviation from the
universality of the scalar masses once we impose the experimental bound
on  $BR(\mu\to e \gamma)$ (light dotted sector inside of the grey
area  in Fig. 2). However
we found that the corresponding limit on  $\tau\to \mu \gamma$ does not
constraint the space of parameters shown in Fig. 3. The fact that the
branching ratios decrease with $m_{3/2}$ is reflected in a wider
light dotted area on the graphic corresponding to $m_{3/2}=400 \ \rm{GeV} $
in Fig.~2.

\section{Conclusions}
We have studied the predictions for the LFV decays $\mu \to e
\gamma$ and  $\tau \to \mu \gamma$ arising from
non universal soft terms. We also showed the dependence of these predictions
on a general, non--diagonal,  texture of the lepton ukawa couplings.

We found the  non--universality of the soft masses is more
relevant for LFV than those of the $A$-term are. However, the latter 
ones are of phenomenological interest for other processes such as
CP violation effects.



\end{document}